 \documentclass[12pt,preprint]{aastex}

\slugcomment{}

\shorttitle{Galaxy number counts in the HDF-N}
\shortauthors{Mar\' \i n-Franch, A}

\begin{document}

\title{Beyond the Hubble Deep Field limiting magnitude: faint galaxy number counts from surface-brightness fluctuations}

\author{A. Mar\' \i n-Franch}
\affil{Instituto de Astrof\' \i sica de Canarias, E-38205 La Laguna,
Tenerife, Spain}
\email{amarin@ll.iac.es}

\and

\author{A. Aparicio}
\affil{Instituto de Astrof\' \i sica de Canarias, E-38205 La Laguna,
Tenerife, Spain}
\email{aaj@ll.iac.es}

\begin{abstract}

The faint end of the differential galaxy number counts, $n(m)$, in the
Hubble Deep Field (HDF) North has been determined for the  F450W, F606W, and F814W filters by means of surface-brightness fluctuation (SBF) measurements. This technique allows us to explore $n(m)$ beyond the limiting magnitude of the HDF, providing new, stronger constraints on the faint end of $n(m)$. This has allowed us to test the validity of previous number count studies and to produce a new determination of the faint end of $n(m)$ for magnitudes fainter than $28.8$  in the $AB$ system and to extend this estimate down to $31$. This value represents an extension of more than two magnitudes beyond the limits of  previous photometric studies. The obtained $n(m)$ slopes are $\gamma=0.27$, $0.21$, and $0.26$ in $B_{450}$, $V_{606}$, and $I_{814}$, respectively.

\end{abstract}


\section{Introduction}

The galaxy luminosity function ($\Phi$) remains at the core  of both galaxy
evolution and cosmology. By integrating $\Phi$ over space and time, various
observable distributions can be obtained. In particular, differential number
counts of galaxies as a function of apparent magnitude, $n(m)$, is obtained by
integrating $\Phi$ over all redshifts and morphological types. The reliability of the predicted $n(m)$ values depends directly on the cosmological model
and on how well the evolution of galaxies is modelled from their formation to the
present. The study of number counts can therefore be used to test world
models or to search for evolution during the look-back time.

The function $n(m)$ provides one of the most fundamental observables and has
been studied by several authors (for reviews see Ellis 1997; Koo \& Kron 1992; Sandage 1988). Recently, a major effort has been made to reach very faint
magnitudes using the {\it Hubble Space Telescope} ({\em HST\/}). Although the {\em HST\/} cannot compete with ground-based observations in terms of collecting area, it does provide an unprecedent view of the optical sky at small angular scales and faint flux levels. A number of authors have studied faint galaxy counts based on the {Hubble Deep Field} (HDF) \citep{W96} and using different
photometry packages \citep{W96,Met96,Lan96,Poz98,Met01}. In this paper, we use the notation $B_{450}$, $V_{606}$, and $I_{814}$ to denote magnitudes in the {\em HST\/} passbands on the $AB$ system \citep{O74}. \citet{Fer98} has
compared the HDF catalogues available in the literature. They generally
show good agreement, although the effects of different isophotal
thresholds and different splitting algorithms are apparent. However,
\citet{Fer00} showed that, at $I_{814}=26$, the different galaxy counts agree
to within $25\%$ in all the catalogues, whereas at $I_{814}=28$ there is a
factor of 1.7 difference among them. This emphasizes the fact that galaxy
counting is not a precise science.

In this paper, the HDF will be used to test the validity of previous number count  studies and to produce a new determination of the faint end of $n(m)$ for magnitudes fainter than $28.8$. Surface-brightness fluctuation (SBF) measurements are used. This allows us to explore $n(m)$ beyond the limiting magnitude of the HDF and to overcome most of the limitations arising from incompleteness, providing new, stronger constraints on the faint end of $n(m)$.

\section{The data}

This work is based on data from the HDF North. Observations were made in 1995 December 18--30, and both raw and reduced data have been put into public domain as a community service \citep{W96}. Version 2 F450W, F606W and F814W images, released into public domain on 1996 February 29, have been used here.

The final Version 2 images were combined using the DRIZZLE algorithm. Drizzling causes the noise in one pixel to be correlated with the noise in the adjacent one \citep{W96}. The SBF technique is based on the spectral analysis of an image. Because the power spectrum of the image is modified by the drizzling process, drizzled images were not considered in this study.
 ``Weighted and cosmic-ray cleaned'' images have been used instead. These are flat-fielded, cosmic-ray-rejected and sky-subtracted stacked images at each dither position. Only dark exposures have been selected. Each of the the three wide-field and the planetary camera WFPC2 chips were analyzed separately. The labels of all  images considered, the corresponding filter and the total exposure times are listed in table \ref{t-data}.

\section{Surface-Brightness fluctuations in the HDF-N}

The SBF  concept was introduced by \citet{TS88}, who noted that, in the
surface photometry of a galaxy far enough away to remain unresolved, a
pixel-to-pixel fluctuation is observed because of the Poisson statistics of the
spatial distribution of stars, globular clusters, background galaxies, etc.
This technique was introduced with the aim of measuring distances. Comparing
SBFs produced by the stellar population of a galaxy with those of nearby
galaxies for which externally calibrated distances are available, accurate
estimates of distances can be obtained up to $\sim$40 Mpc \citep{T00,T01}.
Other authors have used SBF studies to determine the age and metallicity of
unresolved stellar populations of nearby galaxies \citep{Liu00,Vaz01,HMA02};
however, the SBF signal can provide information about other kinds of undetected
and unresolved objects in an image. This is the case, for example, of
globular cluster populations \citep{BT95,B99,MA02a,MA02b}. In this paper, SBFs
have been used to characterize the faint end of $n(m)$ using undetected
galaxies in the HDF-N images.

Next in this section, the  theoretical background of SBFs and the HDF-N signal measurement are described in detail.

\subsection{Theory}

The SBF technique involves spectral analysis of the pixel-to-pixel fluctuation signal. This provides the total point spread function PSF-convolved variance ($P_0$) produced by all point objects whose spatial flux distribution is convolved with the PSF, and the total non-PSF-convolved variance ($P_1$). A convolution in the real space transforms into a product in  Fourier space. For this reason, the power spectrum of an image, $P(k)$, has the form:

\begin{equation} \label{pow1}
P(k)=P_1+P_0 E(k),
\end{equation}
where $E(k)$ is the power spectrum of the PSF convolved with some window function. The $E(k)$ can be approximated by the power spectrum of the PSF alone, $P_{\rm PSF}(k)$, with a negligible error \citep{JTL98}. So
eq.~\ref{pow1} transforms into

\begin{equation} \label{pow2}
P(k)=P_1+P_0 P_{\rm PSF}(k).
\end{equation}

Once the power spectrum of an image has been computed, the variances $P_0$ and $P_1$ can be obtained fitting eq.~\ref{pow2} to data.

In HDF images, $P_0$ is mainly produced by faint galaxies, and, as we will see later, by cosmic rays. So $P_0$ must be equal to the sum of the variances produced by faint galaxies ($\sigma^{2}_{\rm BG}$) and by cosmic rays ($\sigma^{2}_{\rm cr}$):
\begin{equation} \label{P0}
P_0= \sigma^{2}_{\rm BG}+\sigma^{2}_{\rm cr}.
\end{equation}
On the other hand, $P_1$ is the sum of read-out noise ($\sigma^{2}_{\rm ro}$), photon shot noise ($\sigma^{2}_{\rm ph}$), and dark current ($\sigma^{2}_{\rm dc}$) variances:

\begin{equation} \label{P1}
P_1=\sigma^{2}_{\rm ro}+\sigma^{2}_{\rm ph}+\sigma^{2}_{\rm dc}.
\end{equation}

The pixel-to-pixel variance produced by a class of objects can be evaluated
as the second moment of the differential number counts of that object
population \citep{TS88}. In the present study, the target population is
composed by faint undetected galaxies. The brightest individuals, which are
detected, must be masked out before the SBF analysis. If all sources
brighter than a limiting flux ($f_{\rm lim}$) are masked, then the variance
produced by the remaining non-masked population is:
\begin{equation} \label{sigma}
\sigma^{2}_{\rm BG}=\int_0^{f_{\rm lim}} n(f) f^2 df.
\end{equation}

We can put this equation in terms of magnitudes via the relationship
\begin{equation}
m=-2.5 \log(f) + m^*_1,
\end{equation}
$f$ being the flux ($DN$ s$^{-1}$ pix$^{-1}$), and $m^{*}_{1}$ the magnitude of an object yielding one unit of flux per unit time; that is, the photometric zero point listed in Table \ref{t-zero}.

If $n(m)$ is known, then the theoretical variance produced by faint galaxies
can be estimated. Assuming the following pure power-law form for $n(m)$:
\begin{equation} \label{bglf}
n(m)=A_0 10^{\gamma m},
\end{equation}
where $A_0$ is a normalizing constant and $\gamma$ is the slope of the magnitude distribution, the variance produced by the non-masked galaxy population is
then
\begin{equation} \label{sigma_bg}
\sigma_{\rm BG}^2=\frac{A_0}{\ln10(0.8-\gamma)}10^{\gamma m_{\rm c}}10^{0.8(m_1^*-m_{\rm c})},
\end{equation}
where $m_{\rm c}$ is the limiting magnitude; i.e. the magnitude
corresponding to the limiting flux, $f_{\rm lim}$. We will refer to this
variance as the $n(m)$-estimated $\sigma_{\rm BG}^2$ from here on. As we have
shown, in order to compute it, an initial $n(m)$, obtained from a given
photometric catalogue, must be assumed.

\subsubsection{Photometric catalogues}\label{obs}

A number of authors have studied $n(m)$ in the HDF-N. The main discrepancies between previous number count studies occur between \citet{W96} and \citet{Met01}. These coincide for ``bright'' magnitudes, but large differences arise for fainter galaxies,  the \citet{Met01} number counts being larger than those of \citet{W96}.

As pointed out by \citet{Fer98}, there are two reasons for this discrepancy. First, as we go fainter the \citet{Met01} magnitudes become systematically brighter than those of \citet{W96}. \citet{Met01} claimed that the effect of this on the counts is actually quite small, generally $<10\%$. Second, \citet{Met01} find objects that \citet{W96} do not detect at all. This appears to account for the majority of the differences between the data sets. \citet{Met01} argued that virtually all of these objects are merged in the reductions of \citet{W96} but  not in their data.

\begin{figure}
\plotone{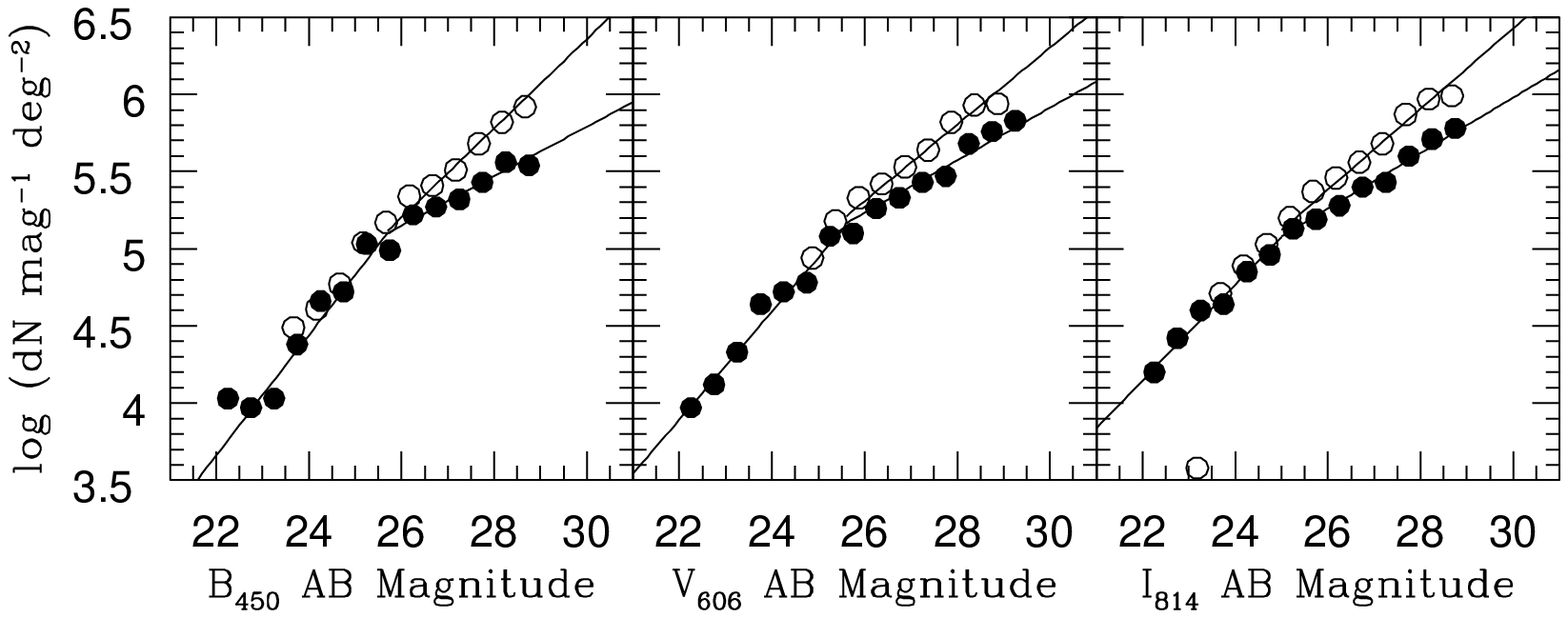}
\caption{Differential number counts of the detected sources by \citet{W96} (filled circles) and \citet{Met01} (open circles) in the F450W, F606W and F814W filters. Solid lines represents the fitted $n(m)$ functions. It can be clearly seen the slope change around magnitude 26 found by \citet{W96}. \label{GLF}}
\end{figure}

In Figure \ref{GLF}, differential number counts results from \citet{W96} (filled circles) and \citet{Met01} (open circles) are plotted for the  F450W, F606W, and F814W filters, respectively. These data have been obtained from tables 9 and 10 in \citet{W96}, and from tables 8, 10 and 12 in \citet{Met01}, all expressed in the total magnitude scale. Following eq.~\ref{bglf}, $n(m)$ has been fitted to the data of \citet{W96} and \citet{Met01}, and the fitted functions (solid lines) are also plotted. As \citet{W96} found a change in the $n(m)$ slope at a magnitude of around 26, $n(m)$ has been fitted to their data in the two magnitude intervals [23, 26] and [26, 29].

The fitted $n(m)$ functions have been used in eq. \ref{sigma_bg} to compute the $n(m)$-estimated $\sigma_{\rm BG}^2$. Results are listed in table \ref{t-sigma}. These $n(m)$-estimated $\sigma_{\rm BG}^2$ values will be later on compared with those directly derived from the SBF measurements (which we will call the SBF-measured $\sigma_{\rm BG}^2$). This comparison will allow us to evaluate the validity of the differential number counts of both \citet{W96} and \citet{Met01} and, as a result, a final $n(m)$ will be proposed.

Before describing the details of the SBF measurements in HDF images, note that the SBF technique is valid only if faint galaxies have a stellar appearance. In \citet{Fer98} the radius--magnitude relation for galaxies in the  \citet{W96} HDF catalogue has been analyzed. For magnitudes fainter than $V_{606}=28.8$, all the galaxies have a radius smaller that $0.16 \arcsec$, close to the FWHM of the WF chips. So very faint galaxies, in the magnitude range where SBF will be measured, can be assumed to have a stellar appearance.

\subsection{Procedure to obtain SBF}

Here, the practical procedure for obtaining the SBF signal in
HDF-N images is described in detail. First of all, it should be
noted that cosmic rays are difficult to discriminate from stars in
{\em HST\/} frames. In order to estimate and eliminate the
cosmic-ray contribution to the SBF signal, $P_0$, a procedure
based on the random nature of  cosmic-ray events has been used for
each filter and chip. The SBF signal has be measured not only on
the final combined images, but on all the individual images listed
in Table \ref{t-data} as well.

Before computing the power spectrum of an image, objects brighter
than $m_{\rm c}$ must be masked out. In this study, the window
functions have been created using the \citet{W96} photometric
catalogue, which is the only one available to us. As isophotal
magnitudes were considered while creating the window functions, in
order to convert them to the total magnitude scale an
isophotal-to-total magnitude correction of 0.2 mag \citep{W96} was
applied. The SBF analysis have been be performed considering two
different values of $m_{\rm c}$: 27.8 and 28.8. All objects
brighter than $m_{\rm c}$ have been masked out using a window
function whose pixel values are zero in a circle centered on the
location of  the bright objects and unity in the rest of the
image. The window function has been created using a patch radius
large enough to completely mask bright galaxies, including their
external haloes and therefore merged galaxies where these exist.
The procedure creating the mask has been the following: first, the
brightest galaxies have been masked one by one manually. In order
to avoid residual light beyond the masked regions, very generous
patch sizes have been adopted. The shape of the used patchs for
these very bright galaxies depends on the shape of the particular
masked galaxy. Once bright galaxies have been masked, the rest of
galaxies brighter than the chosen $m_{\rm c}$ have been masked
using circular patches. The radius of these patches has been
chosen to be the same for all galaxies, and its size is again very
generous: the adopted patch radius is 15 pixels.

In order to test if all the residual light beyond the masked regions has been eliminated, the SBF analysis has been repeated for one image varying the radius of the patches. We have considered patch radii of 20, 25 and 30 pixels. Note that if a radius larger than 30 pixels would be used, the image would be completely masked due to the superposition of adjacent patches. The considered image has been the F450W average image of WF2 with $m_c=27.8$. The SBF results are listed in table \ref{patches}. It can be seen that the SBF results are independent of the patch radius. As a conclusion, it can be seen that the adopted patch radius avoid residual light beyond the masked regions, as required.

If the differences between the \citet{W96} and \citet{Met01} data sets relies on the objects that are merged in the former  and not in the latter, as argued by \citet{Met01}, then the created window function is virtually the same as one would obtain using the \citet{Met01} photometric catalogue, and  also masks objects that are in this catalogue and not in that of \citet{W96}.

Next, multiplying the window function by the image, the residual masked image is obtained. It is this image which is used to compute the power spectrum. As  has been said, two sets of masked images have been computed, one for each $m_c$ value.

The power spectrum of the masked image is two-dimensional. It is radially averaged in order to obtain the one-dimensional power spectrum. Fitting the power spectrum of the images with eq.~\ref{pow2}, the quantities $P_0$ and $P_1$ can be obtained for each image. As we could not construct a PSF from HDF images, since there were not enough stars in any of the four chips, the used PSFs were the high-S/N PSFs extracted by P. B. Stetson from a large set of uncrowded and unsaturated WFPC2 images.

Spatial variations of the PSF along the CCD mosaic could introduce a
significant uncertainty in $P_0$. To limit this effect, we have used a PSF
template for each one of the four chips of the WFPC2. Minor PSF variations
inside each particular chip have not been considered. For each chip,
both PSF and the computed power spectrum represent mean values across
the complete field of the chip, so any spatial variation of the PSF across
the chip would affect the SBF fitting procedure, introducing an uncertainty in the $P_0$ measurement. This uncertainty is small, and in any case, it is included in the obtained $P_0$ uncertainty.

In Fig.~\ref{$B_{450}$.average.29} we show an example of the SBF fitting procedure in an HDF image. The observed discrepancy at low wave numbers between the obtained power spectrum and the fit is due to large scale fluctuations in the background brightness of the images. This wave number region, where the discrepancy occurs, is not taken into account when fitting eq.~\ref{pow2}. The power-spectrum fitting procedure is the following: eq.~\ref{pow2} is used to fit the power spectrum for wave numbers in the range [$k_0$, $k_{\rm max}$],  $k_{\rm max}$ being the highest wave number of the computed power spectrum, and $k_0$ a number which we vary from 0 to $k_{\rm max}$. As a result, two functions, $P_0(k_0)$ and $P_1(k_0)$, are obtained. The function $P_0(k_0)$ is also shown in
Fig.~\ref{$B_{450}$.average.29} (small boxes). It can be seen that this function exhibits a ``plateau'' region. The final adopted result and its uncertainty for $P_0$ are obtained computing the average and standard deviation of $P_0(k_0)$ in the plateau interval. For the $P_1$ measurement, the procedure is exactly the same as for $P_0$.

$P_0$ and $P_1$ results corresponding to $m_c=27.8$ and $m_c=28.8$ are listed in Tables \ref{t-results28} and \ref{t-results29}, respectively. Results are listed for all images, corresponding to all chips and all filters.

\section{Results}

In this section, results from Tables \ref{t-results28} and \ref{t-results29} will first be carefully analyzed, and will be used to test the probable effect of flat-fielding errors on the measured $P_0$. Then, the obtained $P_0$ values in different images will be used to estimate the contribution of cosmic rays and to deduce the desired SBF-measured $\sigma_{\rm BG}^2$. Finally, in order to check the SBF results, two consistency tests will be performed.

\subsection{The effect of flat-fielding errors on $P_0$}

The possibility of flat-fielding errors affecting the measured $P_0$ must considered. If flat-fielding would contribute to $P_0$, then its effect should be larger in images with high sky background. In this context, considering same filter and chip images, if flat-fielding is contributing to $P_0$, a relation between $P_0$ and the sky level should appear.

\begin{figure}
\plotone{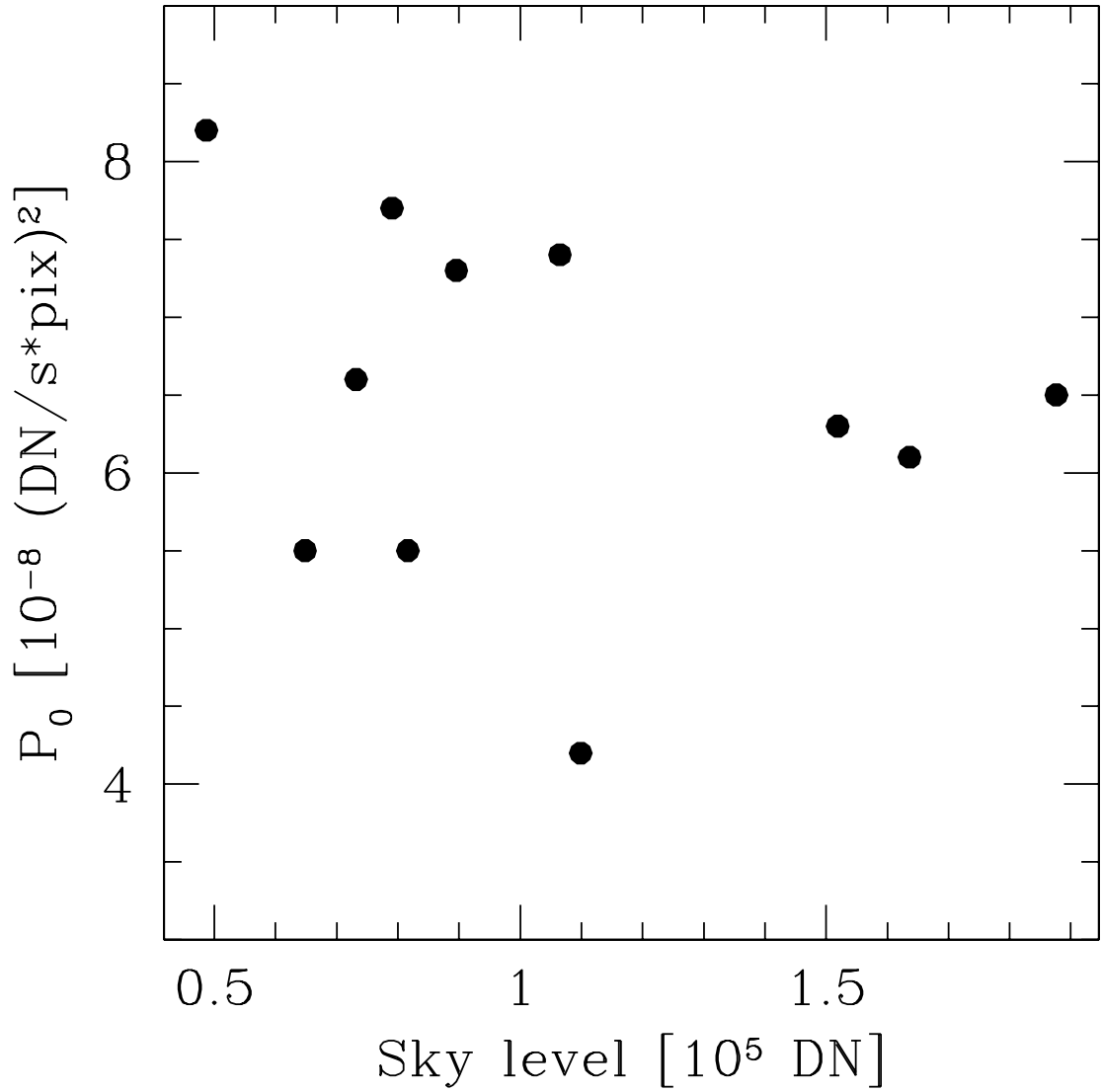}
\caption{$P_0$ results versus the sky level in WF2 F606W images with $m_c=27.8$. The absence of a trend indicates that flat-fielding errors are not significant in the measurement of $P_0$. \label{flatfielding}}
\end{figure}

To test if the flat-fielding errors have an influence on the measured $P_0$, filter F606W has been considered because it provides a larger number of images covering a wider range of sky levels. Sky levels are provided in the weighted, cosmic-ray cleaned image headers. As an example, fig. \ref{flatfielding} shows the measured $P_0$ as a function of these sky levels for the WF2 F606W images with $m_c=27.8$. It can be seen that no relation is apparent, so flat-fielding errors are insignificant in the $P_0$ measurements.

\subsection{The effect of cosmic-rays and SBF-measured $\sigma_{\rm BG}^2$ estimation}

The {\em HST\/} provides images of exceptional resolution. As a consequence, discriminating stars from cosmic rays is a difficult task. This also has implications  for SBF measurements. If a cosmic-ray event alters only one pixel, then it is handled as white noise, and consequently, it contributes to the $P_1$ signal. But if two or more pixels are affected, then a cosmic ray can be confused with a point source and  it will therefore contribute to the $P_0$ signal. In the WFPC2, the majority of cosmic ray events affect 2--3 pixels, so if undetected cosmic rays exist in an image, the SBF results will be affected by an increase of both $P_1$ and $P_0$. But there is noise also due
to detected cosmic rays. The pixels containing cosmic rays have been masked during the data reduction of the images, and therefore, the noise in these pixels is higher than it is in pixels where none of the input images were masked.

As we will see later, the contribution of cosmic rays to $P_0$ is not negligible, and is necessary to control this effect in order to obtain the SBF-measured $\sigma_{\rm BG}^2$. It is for this reason that a strategy aimed at estimating and eliminating the contribution of cosmic rays from $P_0$ must be developed. Such a strategy is described next.

In an individual weighted and cosmic-ray cleaned image, $P_0$ has two contributions, faint galaxies and cosmic rays:
        \begin{equation}\label{cr1}
        P_0=\sigma_{\rm BG}^2+\sigma_{\rm cr}^2.
        \end{equation}

If such individual images averaged, the contributions from galaxies and cosmic rays will be different, because of their different natures. Measured in the averaged image, $P_0$ becomes
        \begin{equation}\label{cr2}
        P_{0}^{\langle {\rm av}>}=\sigma_{\rm BG}^2+\frac{\sigma_{\rm cr}^2}{N},
        \end{equation}
where $N$ is the number of individual images used for the average.

In Tables \ref{t-results28} and \ref{t-results29}, results for $P_0$  are listed
for all the individual weighted and cosmic-ray cleaned images and for the averaged images as well. If the influence of cosmic rays were negligible, $P_0$ would be the same in all the individual images and in the averaged one. It can be seen that this is not the case, from which we may conclude that the contribution of cosmic rays to $P_0$ is  non-negligible.

Equations \ref{cr1} and \ref{cr2} can now be used to obtain $\sigma_{\rm BG}^2$
and $\sigma_{\rm cr}^2$. For $P_0$, the mean of the values of the individual
images for each chip and filter have been used. Results are listed in Table
\ref{t-results-bg} for the two considered values of $m_c$. The final
SBF-measured $\sigma_{\rm BG}^2$
values for each filter can now be  obtained from the
averages of the single-chip results. They are given in the last row of each
$m_c$ set in Table \ref{t-results-bg} and will be compared in \S
\ref{discussion} with the $n(m)$-estimated $\sigma_{\rm BG}^2$ values.

\subsection{Two consistency tests}

In this section, a couple of consistency tests have been performed
 in order to check the reliability of our SBF-measured $\sigma_{\rm BG}^2$ results.

\subsubsubsection{Comparison of expected and measured $P_1$}

Together with the PSF-convolved variance, $P_0$, SBFs provide the value of $P_1$, the non-PSF-convolved variance. This can be compared with its expected value, directly obtained from the read-out noise, the dark current, and the sky brightness analysis of each image.

For HDF Version 2, each of the weighted and cosmic-ray cleaned images is the result of combining several exposures with the same dither position. The different exposures are combined with weights proportional to the inverse variance ($1/P_1$) at the mean background level. The variance $P_1$, in electrons, is computed from the following noise model \citep{W96}:

\begin{equation}
P_1=bt+dt+r^2,
\end{equation}
where $t$ is the exposure time, $b$ is the sky background rate, $d$ is the
dark current, and $r$ is the read-out noise. The inverse variances, $1/P_1$,
of each exposure are provided in the header of the resulting weighted and
cosmic-ray cleaned image. From this information, the value of $P_1$
corresponding to the latter can be computed. As an example, these values are
listed in Table \ref{t-p1} (column 2) for the WF2 images and $B_{450}$. The
$P_1$ values obtained directly from the SBF analysis of the images, using
$m_{\rm c}=28.8$, are listed in column 3. Both computed and observed values of
$P_1$ are equivalent in all cases. Only a slight excess in the observed $P_1$
is noticeable. This excess is produced by cosmic rays, which also
contribute to the measured $P_0$ values, as we have shown.

This test shows that, with this technique, the white noise ($P_1$) is
determined with high precision, thereby reinforcing the correctness of $P_0$
measurements.

\subsubsubsection{Comparison of estimated and measured $[\sigma_{\rm BG}^2]$}\label{interval}

This second consistency test is based on the comparison of the parameter $[\sigma_{\rm BG}^2]$ computed in two different ways. We call $[\sigma_{\rm BG}^2]$ to the variance produced by galaxies with magnitudes within a given interval:

\begin{equation}
[\sigma_{\rm BG}^2](m_{A},m_{B}) \equiv \int_{f_{B}}^{f_{A}} n(f) f^2 df,
\end{equation}
where $m_{A}$ is the magnitude corresponding to a flux $f_{A}$, and $m_{B}$
to $f_{B}$. Defined in this way, $[\sigma_{\rm BG}^2]$ is the difference between
the variances computed with two values of $m_{\rm c}$, namely $m_{\rm
  c}=m_{A}$ and $m_{\rm c}=m_{B}$:

\begin{eqnarray}
[\sigma_{\rm BG}^2](m_{A},m_{B})=\sigma_{\rm BG}^2(m_{\rm c}=m_{A})-\sigma_{\rm BG}^2(m_{\rm c}=m_{B})
\nonumber\\
\end{eqnarray}

The $n(m)$-estimated and SBF-measured $[\sigma_{\rm BG}^2]$ can be now
compared. To do this, the magnitude interval [27.8, 28.8] has been considered.

As the \citet{W96} photometric catalogue has been used to create the window
functions, only objects found by them in the magnitude interval [27.8, 28.8]
contribute to the SBF-measured $[\sigma_{\rm BG}^2]$. In particular, some of
the objects within the interval [27.8, 28.8] in the \citet{Met01} catalogue will
remain masked, namely those which, following \citet{Met01}, are merged in the
\citet{W96} catalogue. This implies that the SBF-measured $[\sigma_{\rm BG}^2]$
is not expected to coincide with the $n(m)$-estimated $[\sigma_{\rm BG}^2]$
computed using the \citet{Met01} data. On the other hand, the
$n(m)$-estimated $[\sigma_{\rm BG}^2]$ computed using the \citet{W96} data
should, for rigor, be similar to the SBF-measured $[\sigma_{\rm BG}^2]$ only if
all the objects of the \citet{W96} catalogue in the interval [27.8, 28.8] are
unmerged. However, if merged galaxies exist in other intervals
\citep{Met01}, there is no reason why they should not be present here also.

The effect of mergers on $[\sigma_{\rm BG}^2]$ can be tested considering that a
fraction of \citet{W96} objects in the interval [27.8, 28.8] are the result of a
merger between two fainter galaxies with integrated fluxes $f_1$ and $f_2$.
We have considered three simple situations: i) $f_1=f_2$; ii) $f_1=2f_2$;
and iii) $f_1=3f_2$. It should be noted that since SBFs are a measure of the
second moment of the brightness function, the more similar  $f_1$ and
$f_2$ are, the larger is the effect introduced in $[\sigma_{\rm BG}^2]$. So case i)
is the most pessimistic and, although it is unrealistic,  will give the maximum
expected effect on $[\sigma_{\rm BG}^2]$ for a given fraction of mergers.

\begin{figure}
\plotone{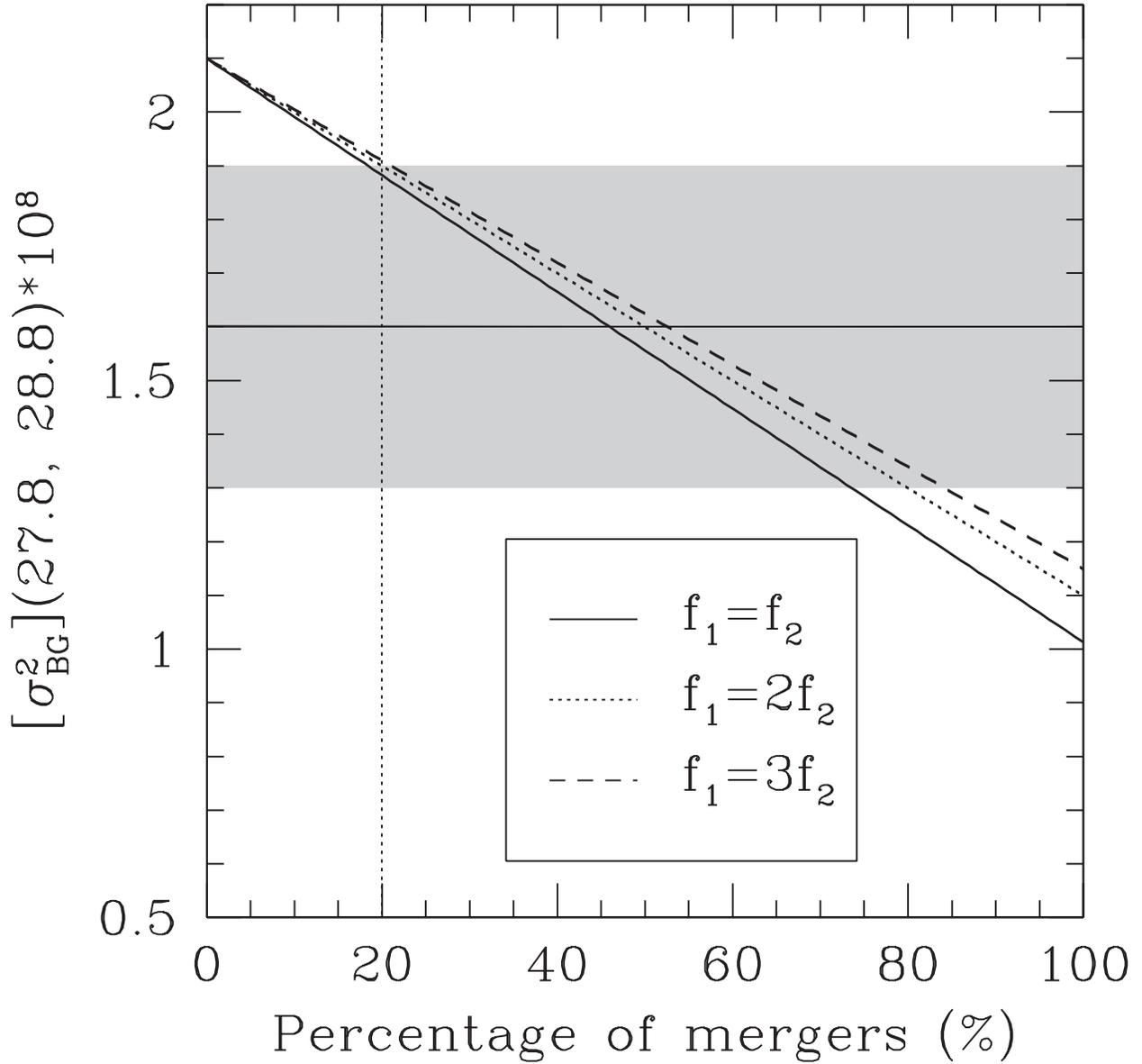}
\caption{Effect on $[\sigma^2_{BG}]$ of a percentage of mergers in the \citet{W96} photometric catalogue in the  F606W filter. Three simple cases have been considered. i) solid line: $f_1=f_2$; ii) dashed line: $f_1=2f_2$; and  iii) short-dashed line: $f_1=3f_2$,  $f_1$ and $f_2$  being the fluxes of the mergered galaxies. The gray region represents the obtained $\pm1\sigma$ SBF-measured $[\sigma^2_{BG}]$. \label{mergers}}
\end{figure}

In Figure \ref{mergers} we show the results of $n(m)$-estimated
$[\sigma_{\rm BG}^2]$ for the three cases and the F606W filter, considering
different values of the percentage of merged objects. It can be seen that,
even in the very pessimistic case where $50\%$ of the \citet{W96} objects
are mergers of two identical galaxies, their influence on
$[\sigma_{\rm BG}^2]$ is less than $25\%$. For example, in a perhaps more realistic
situation in which about 20--25\% of the objects are mergers of two galaxies
with different magnitudes, the effect on $[\sigma_{\rm BG}^2]$ would smaller than
$\sim$10\% (see figure \ref{mergers}). In conclusion, the effect of mergers
on $[\sigma_{\rm BG}^2]$ computed using the \citet{W96} number
counts is small and can be expected to remain below about 15\% for any
reasonable scenario.

We can hence proceed with our test on $[\sigma_{\rm BG}^2]$ for the [27.8, 28.8]
interval using the \citet{W96} data. The results for SBF-measured and
$n(m)$-estimated $[\sigma_{\rm BG}^2]$ are given in Table \ref{t-interval}. The
value of $n(m)$-estimated $[\sigma_{\rm BG}^2]$ has been computed using
\citet{W96} and \citet{Met01} data. The results are given in columns 2 and 3,
while column 4 lists the $n(m)$-measured $[\sigma_{\rm BG}^2]$ values. It can be seen that they are not compatible with the values obtained from the \citet{Met01} data, as expected.

On the other hand, the $n(m)$-estimated $[\sigma_{\rm BG}^2]$ obtained from the \citet{W96} number counts and the SBF-measured  $[\sigma_{\rm BG}^2]$ are very close. In order to compare them with more detail, let analyze figure \ref{mergers} again. The $n(m)$-measured $[\sigma_{\rm BG}^2]$ value for the F606W filter and its error interval have also been plotted in the figure (shadowed region). It can be seen now that they fully coincide if a number of mergers about 20--30\% is assumed. This situation is realistic and compatible with \citet{Met01}'s claims.

Summarizing, the test has been successful and shows that the SBF measurements are well-calibrated.

\section{Discussion}\label{discussion}

In this section, SBF-measured $\sigma_{\rm BG}^2$ results (listed in Table
\ref{t-results-bg}) and the $n(m)$-estimated $\sigma_{\rm BG}^2$ values obtained from both the
\citet{W96} and \citet{Met01} data (listed in Table \ref{t-sigma}) will be compared. In the following,
two possibilities will be considered and their consequences discussed:
i) that the \citet{W96} data represent the right differential number counts, and
ii) that the \citet{Met01} number counts
 are correct. The obtained SBF measurements will be used to test the validity of these
  possibilities and, as a result, a final $n(m)$ will be proposed.

\subsection{Option A: Assuming \citet{W96} galaxy number counts}

Here we assume that the
\citet{W96} number counts are correct. Comparing the $n(m)$-estimated $\sigma_{\rm BG}^2$
 obtained using \citet{W96} data
  and SBF-measured $\sigma_{\rm BG}^2$, it can be seen that the former are much larger.
  There are only two possible sources to account for this discrepancy: first, a faint
  unresolved stellar population, belonging to the Milky Way halo, could be responsible
  of the excess in the SBF signal; and second, the faint end of $n(m)$ is different from
  the fitted one used here to evaluate the $n(m)$-estimated $\sigma_{\rm BG}^2$. In the
  first case, SBF results may be used to characterize such a halo population. In the second
  case, SBF results may be used to deduce a new faint end of $n(m)$ able to account for
  the SBF-measured $\sigma_{\rm BG}^2$.

We shall analyze both possibilities in detail and discuss the feasibility of each one and its
 compatibility with the observations; finally, we shall deduce its implications for $n(m)$.

\subsubsection{Faint Milky Way halo stars}

Here, we will consider  a Milky Way halo population of faint stars  be responsible of the observed excess in the SBF signal. We will firstly deduce the halo population necessary to cause this SBF signal excess. In order to check the feasibility of this hypothesis, the obtained halo population will  then
be compared with observations in the HDF.

Lets consider a simple population of objects with absolute magnitude $M$
following the standard spatial distribution used by \citet{BT87}:
        \begin{equation}\label{eq1}
        \rho(r)dr=\frac{\rho_0}{1+(r/a)^{\alpha}}dr,
        \end{equation}
where $\rho(r)dr$ is the number of objects per pc$^3$ at a distance from the Milky Way center between $r$ and $r+dr$; $a$ is the core radius, and $\rho_0$ is the object density in the Milky Way center. For simplicity we take $\alpha=2$.

To derive the SBF signal from the former population, we must first express the equations in terms of  distance from the Sun ($\delta$). This can be done using:
        \begin{equation}
        r^2=\delta^2+r_0^2-2 \delta r_0 \cos b \cos l,
        \end{equation}
where $r_0$ is the galactocentric radius, i.e., the distance from the Sun to the Milky
Way center, and $(b,l)$ are  galactic coordinates. The spatial
distribution of objects expressed in  spherical coordinates is then:
        \begin{eqnarray}
        \lefteqn{ n(\delta,\theta,\varphi)dV  {}}
        \nonumber\\
        & & {}=\frac{\rho_0 a^2}{a^2 + r_0^2 + \delta^2 - 2 r_0 \delta \sin \theta \cos \varphi} \delta^2 \sin \theta d \delta d \theta d \varphi .
        \end{eqnarray}

Integrating for the HDF-N and considering its coordinates $l=127 \arcdeg$ and $b=54 \arcdeg$, the
former expression is reduced to:
        \begin{equation}
        n(\delta)d \delta \simeq \frac{1.38 \times 10^{-13} \rho_0 a^2 \delta^2}{a^2 + r_0^2 + \delta^2 - 0.7075 r_0 \delta}d \delta .
        \end{equation}

Now, $n(\delta)d \delta$ can be written in terms of magnitudes by means of the distance modulus to obtain:
        \begin{eqnarray}\label{eq2}
        \lefteqn{ n(m)d m  {}}
        \nonumber\\
        & & {}= \frac{6.36 \times 10^{-11} \rho_0 a^2 10^{0.6(m-M)}}{a^2 + r_0^2 + 100 \times 10^{0.4(m-M)}+7.075 r_0 10^{0.2(m-M)}} dm .
        \end{eqnarray}

The number of resolved objects with $m<28.8$ that should appear in the HDF-N,
($N_{\rm HDF-N}^{(m<28.8)}$) can be now deduced from eq.~\ref{eq2}, as well as the variance that the faint part of the population ($m>28.8$) would produce ($\sigma_{\rm HDF-N}^{2(m>28.8)}$):
        \begin{eqnarray}\label{eq3}
        \lefteqn{ N_{\rm HDF-N}^{(m<28.8)}=1.31 \times 10^{-4} \rho_0 a^2 {}}
        \nonumber\\
        & & {}\times \int_{-\infty}^{28.8-M}\frac{10^{0.6x}}{a^2+r_0^2+100 \times 10^{0.4x}+7.075 r_0 10^{0.2x}}dx   ,
        \end{eqnarray}

                \begin{eqnarray}\label{eq4}
        \lefteqn{ \sigma_{\rm HDF-N}^{2(m>28.8)}=6.36 \times 10^{-11} \rho_0 a^2 10^{0.8(m_1^*-M)} {}}
        \nonumber\\
        & & {}\times \int_{28.8-M}^{\infty}\frac{10^{-0.2x}}{a^2+r_0^2+100 \times 10^{0.4x}+7.075 r_0 10^{0.2x}}dx   .
        \end{eqnarray}

\begin{figure}
\plotone{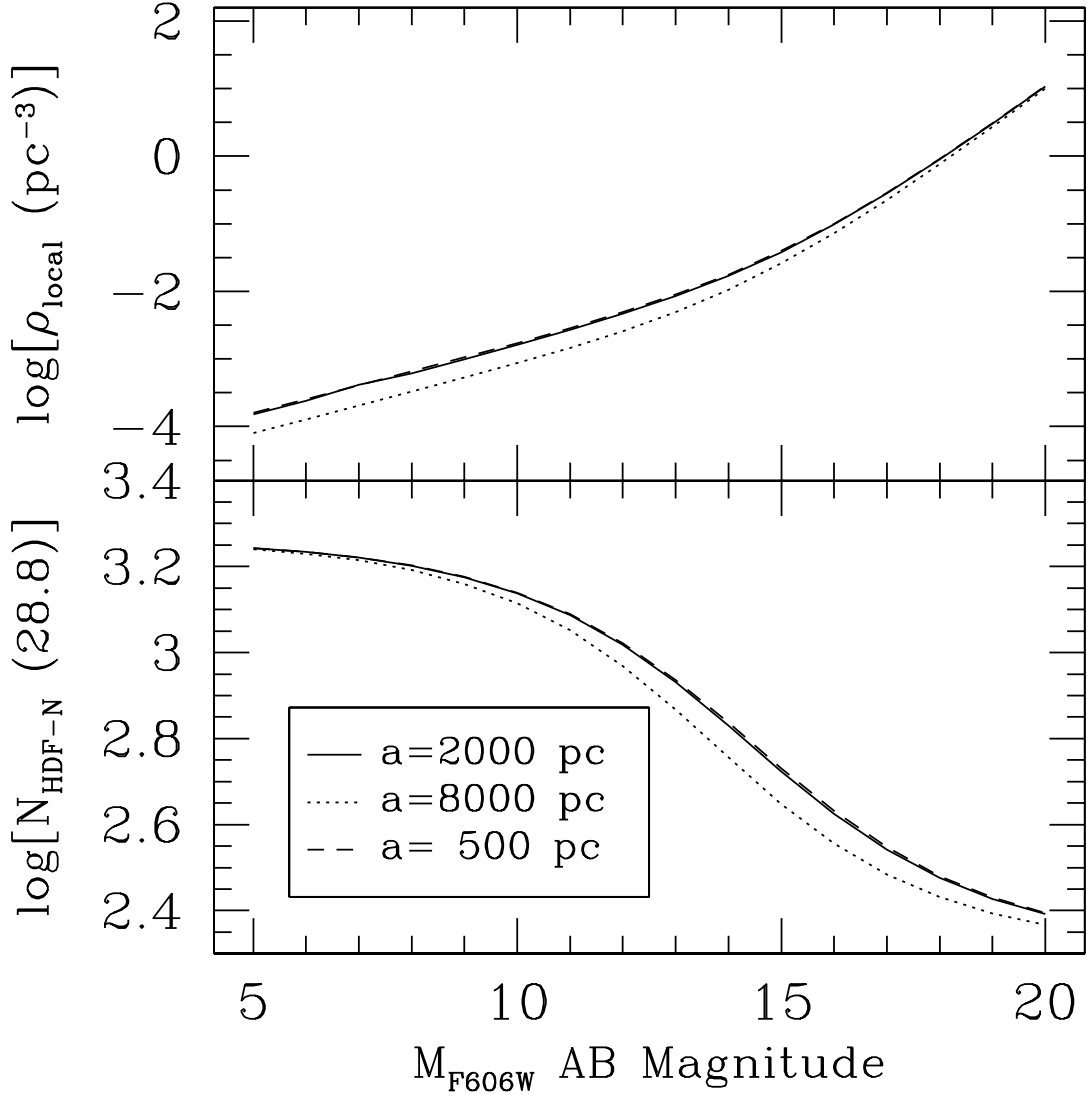}
\caption{Local density ($\rho_{\rm local}$) and number of objects in the HDF-N images up to
magnitude $V_{606}=28.8$ ($N_{\rm HDF-N}(28.8)$) predicted by the considered halo model for three
possible values of core radius, $a$. These results have been obtained assuming that the
excess in $P_0$ is completely produced  by halo objects. It can be seen that the model
predictions are not compatible with observations. This implies that the observed $P_0$
excess cannot be caused by halo objects. See text for details. \label{mod_halo}}
\end{figure}

In order to compare these predictions with the HDF data, the $V_{606}$ filter
results have been used. A value must be assumed for the core radius, $a$.
Realistic values are around $2000$ pc \citep{BS80}, but $500$ pc and $8000$
pc have been also used to check a wide interval of possibilities. Making
$\sigma_{\rm HDF-N}^{2(m>28.8)}$ equal to the $\sigma_{\rm BG}^{2}$ excess observed in
the filter $V_{606}$ for $m_c=28.8$ (that is, $4.87 \times 10^{-9}$ $\left[DN/
({\rm s\ pix})\right]^2$, see Tables \ref{t-sigma} and \ref{t-results-bg}) and
introducing the value in eq.~\ref{eq4}, the central density $\rho_0$ can
be obtained and used in eqs
~\ref{eq1} and \ref{eq2} to derive
$\rho_{\rm local}$ and $N_{\rm  HDF-N}^{(m<28.8)}$. Results are plotted in
Fig.~\ref{mod_halo}, where $\rho_{\rm local}$ and $N_{\rm HDF-N}^{(m<28.8)}$ are shown
as functions of $M_{\rm F606W}$, the absolute magnitude of the halo population
objects.

This figure implies the existence of a large number of halo objects that
should be present in the HDF images. Note that $N_{\rm HDF-N}^{(m<28.8)}$ is greater
than $\sim$250 in all cases. This result is not compatible with the HDF-N
observations, where no obvious stars are present, except for a few 20th
magnitude ones, \citep{K96,F96}.

In conclusion, the observed excess in SBF-measured $\sigma_{\rm BG}^{2}$
cannot be produced by objects belonging to the Milky Way halo. Otherwise a large
number of resolved objects from this halo population would show up  in the
HDF-N images, which is not the case.

\subsubsection{Faint galaxy number counts}

If the observed $\sigma_{BG}^{2}$ excess cannot be produced by Milky Way
halo objects, the only possibility is that it is caused by faint galaxies. The large excess obtained in the
SBF-measured $\sigma_{BG}^{2}$ with respect to the $n(m)$-estimated $\sigma_{BG}^{2}$ would
imply an increase in the slope of $n(m)$ at some magnitude fainter than $m_c=28.8$. This slope can be
computed by fitting the $n(m)$-estimated $\sigma_{BG}^{2}$ to our SBF-measured $\sigma_{BG}^{2}$ and
taking the slope as a free parameter.
If it is assumed that the slope change
 occurs at $m_c = 28.8$ for all filters, the
  resulting slopes for the fainter range are $\gamma=0.60$, $0.44$, and $0.54$ for
  $B_{450}$, $V_{606}$, and $I_{814}$, respectively. These slopes would be valid
  up to $B_{450}=34.4$, $V_{606}=31.9$, and $I_{814}=32.5$ at least since the
  contribution of fainter magnitudes to $\sigma_{BG}^{2}$ becomes smaller than
  the uncertainties in the SBF-measured $\sigma_{BG}^{2}$ results. If the slope
  change were to occur at a magnitude fainter than 28.8, it would result in a steeper
  $n(m)$. In any case, such  big changes in the slope of $n(m)$  seem unrealistic. In
  our opinion, this possibility should be rejected. As a consequence, it must be
  concluded that the \citet{W96} data are incomplete.

\subsection{Option B: Assuming \citet{Met01} galaxy number counts}

Assuming that the \citet{Met01} differential number counts are correct, the SBF-measured $\sigma_{\rm BG}^2$ results
 listed in Table \ref{t-results-bg} and the $n(m)$-estimated $\sigma_{\rm BG}^2$ values obtained using
 the  \citet{Met01} data, listed in Table \ref{t-sigma}, can be compared.

It can be seen that the SBF-measured and $n(m)$-estimated $\sigma_{\rm BG}^2$  coincide within the error bars for the F814W filter, and is very similar for the F450W filter. Only in the filter F606W some differences arise. This implies that
 extrapolation of the \citet{Met01} $n(m)$ function to magnitudes fainter than 28.8  accounts
 almost entirely for the measured SBF signal, thus
 indicating a high level of precision in the \citet{Met01} data.

However, as SBF-measured and $n(m)$-estimated $\sigma_{\rm BG}^2$ present
slight differences in the F450W and F606W filters, the most likely $n(m)$ function that completely fits
 our SBF measurements can be determined. This will be done in the next subsection.

\subsection{The galaxy differential number counts beyond $V_{606}=28.8$}

In this section, the most likely $n(m)$ function will be obtained for magnitudes fainter than 28.8. We consider that the most likely $n(m)$ function is that which fits our SBF measurements.

\begin{figure}
\plotone{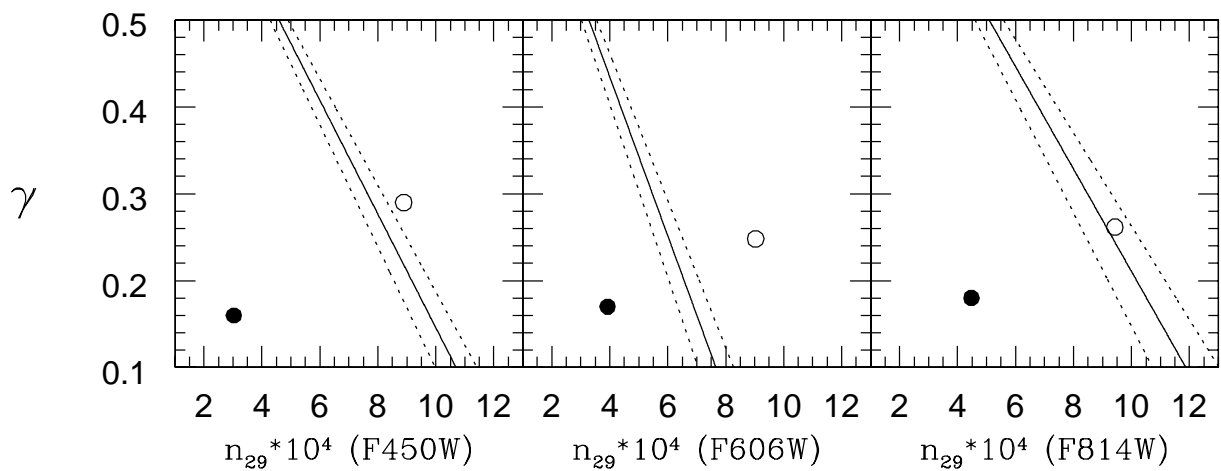}
\caption{The $n(m)$ slope, $\gamma$, versus the number of galaxies in the HDF-N in the magnitude interval
[28.75, 29.25], $n_{29}$, for the  F450W, F606W, and F814W filters. Solid lines:
the pairs ($\gamma$, $n_{29}$) that would produce our measured SBF signal; short-dashed lines:
the pairs ($\gamma$, $n_{29}$) that would produce the SBF measurements
$\pm 1\sigma$; filled circle: the set ($\gamma$, $n_{29}$) corresponding to the
extrapolation of $n(m)$ obtained from the \citet{W96} data; open circle: the same corresponding to
\citet{Met01}. It can be seen that the
 \citet{Met01} number counts nearly fit out SBF measurements. See text for details. \label{SBF}}
\end{figure}

An $n(m)$ function is completely determined by
giving the slope ($\gamma$) and the number of galaxies in a given area and magnitude interval ($n_m$).
In particular we consider $n_{29}$, computed for the magnitude interval [28.75, 29.25] and the area of
the HDF-N. Note that there are infinite sets of ($\gamma$, $n_{29}$) that can produce the same SBF signal.
 In Figure \ref{SBF}, the  ($\gamma$, $n_{29}$) pairs that can account for our measured SBF signal have
 been plotted (solid lines) for the  F450W, F606W, and F814W filters. With short-dashed lines we
  represent the  ($\gamma$, $n_{29}$) pairs that would produce the SBF measurements $\pm 1\sigma$.
  In this figure, the  ($\gamma$, $n_{29}$) sets corresponding to the extrapolation of $n(m)$ obtained from
  \citet{W96} (solid circle) and \citet{Met01} data (open circle) have  also been plotted.

It can be seen in Figure \ref{SBF} that the \citet{Met01} ($\gamma$, $n_{29}$)  reproduce our
SBF results for the  F814W filter, as previously shown. For F450W, the \citet{Met01}
($\gamma$, $n_{29}$) is very close to our SBF measurements. In all cases (except for F606W),
the \citet{W96} ($\gamma$, $n_{29}$) results are far from our SBF results, as previously argued.

The most likely $n(m)$ function can be obtained from Figure \ref{SBF}. We assume that, for each filter, the best $n(m)$ estimate for magnitudes fainter than 28.8 is given by the nearest point of the solid lines to the \citet{Met01} point. In this case, results for the slopes are $\gamma=0.27$, $0.21$, and $0.26$ for $B_{450}$, $V_{606}$, and $I_{814}$, respectively.

\begin{figure}
\plotone{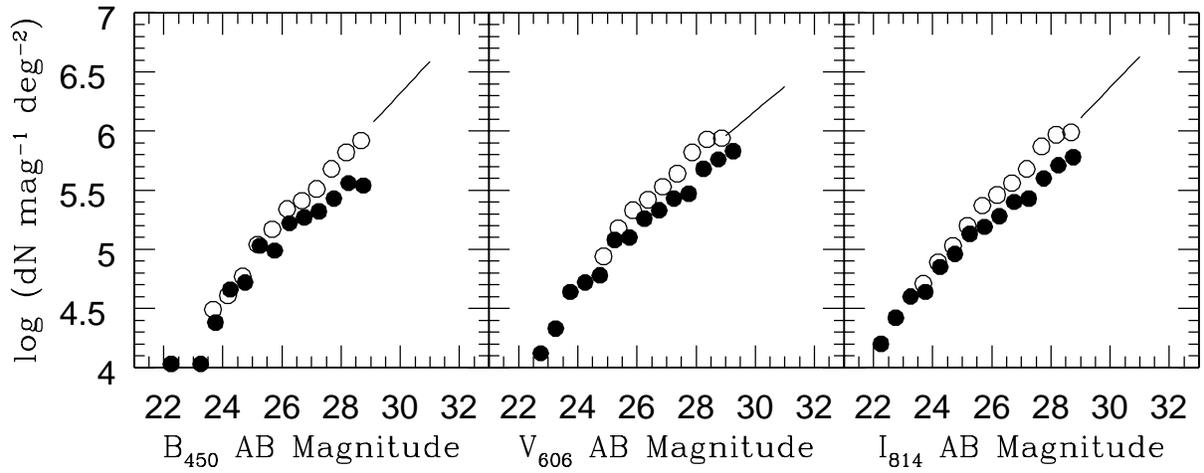}
\caption{$n(m)$ final results in the F450W, F606W and F814W filters. Filled circles represent \citet{W96} differential number counts, open circles are the results from \citet{Met01}. Solid line represents the obtained faint end (fainter than magnitude 28.8) of $n(m)$. The obtained $n(m)$ function is valid down to magnitude 31, at least. \label{GLF2}}
\end{figure}

The results are listed in Table \ref{t-GLF} and plotted in Figure \ref{GLF2} (solid
lines). The  slopes obtained are valid down to magnitude $31$ at least.
The contribution to the SBF signal by objects of fainter magnitudes is less than
the uncertainty in the SBF measurements. This value represents an extension
of more than two magnitudes beyond the limits of the previous photometric
studies by \citet{W96} and \citet{Met01}. It should also be mentioned that,
within the uncertainties and except for the objects that could be merged into
brighter ones in \citet{Met01}, it is free from incompleteness.

\section{Conclusions}

In this paper, the faint end of the differential galaxy number counts,
$n(m)$, has been studied by means of SBF measurements. Once the contribution from cosmic rays has been evaluated and eliminated from the SBF signal, the background PSF-convolved variance
  originating from faint objects has been
carefully analyzed. Our conclusions can be summarized as follows:

\begin{itemize}
\item Comparing the SBF-measured $\sigma_{\rm BG}^2$ with the $n(m)$-estimated
$\sigma_{\rm BG}^2$ predicted by the extrapolation of \citet{W96} number counts
 a clear excess has been found in the measured signal. The possibility that the excess might
  be produced by Milky Way halo stars is ruled out because it would be totally
   incompatible with the resolved stellar population present in the HDF. On the other hand, if this excess is caused
   by a faint galaxy population modifying the faint end of $n(m)$, then the required slopes for magnitudes fainter than
   28.8 are $\gamma=0.60$, $0.44$, and $0.54$ for $B_{450}$,
$V_{606}$, and $I_{814}$, respectively. Such  big changes in the $n(m)$ slope seem unrealistic. In our opinion,
 this possibility should be rejected. In conclusion, the \citet{W96} number counts are not compatible with our
 SBF measurements, probably owing to the incompleteness in their data.

\item Comparing the SBF-measured $\sigma_{\rm BG}^2$ with the $n(m)$-estimated
$\sigma_{\rm BG}^2$, predicted by the extrapolation of \citet{Met01} number counts, we find that
they coincide within the error bars for the  F814W and F450W filters and are similar for F606W.
This implies that the extrapolation of the \citet{Met01} $n(m)$ function to
magnitudes fainter than 28.8 nearly accounts for the measured SBF signal, indicating a high level of precision
in the \citet{Met01} results.

\item The most likely $n(m)$ function has been obtained fitting our SBF
  results. Results for the $n(m)$ slope for magnitudes fainter than $28.8$ are
  $\gamma=0.27$, $0.21$, and $0.26$ for $B_{450}$, $V_{606}$, and $I_{814}$,
  respectively. The obtained slopes
  are valid down to magnitude $31$ at least. This value represents an
  extension of more than two magnitudes beyond the limits of the previous
  photometric studies by \citet{W96} and \citet{Met01}.

\end{itemize}

\acknowledgments

This work is based on observations with the NASA/ESA {\it Hubble Space
Telescope}, obtained in the Space Telescope Science Institute, which is
operated by the Association of Universities for Research in Astronomy, Inc.
(AURA), under NASA contract NAS5-26555. This research has been supported by
the Instituto de Astrof\'\i sica de Canarias (grant P3/94), the DGESIC of the
Kingdom of Spain (grant PI97-1438-C02-01), and the DGUI of the autonomous
government of the Canary Islands (grant PI1999/008).

\clearpage

\clearpage

\begin{figure}
\plotone{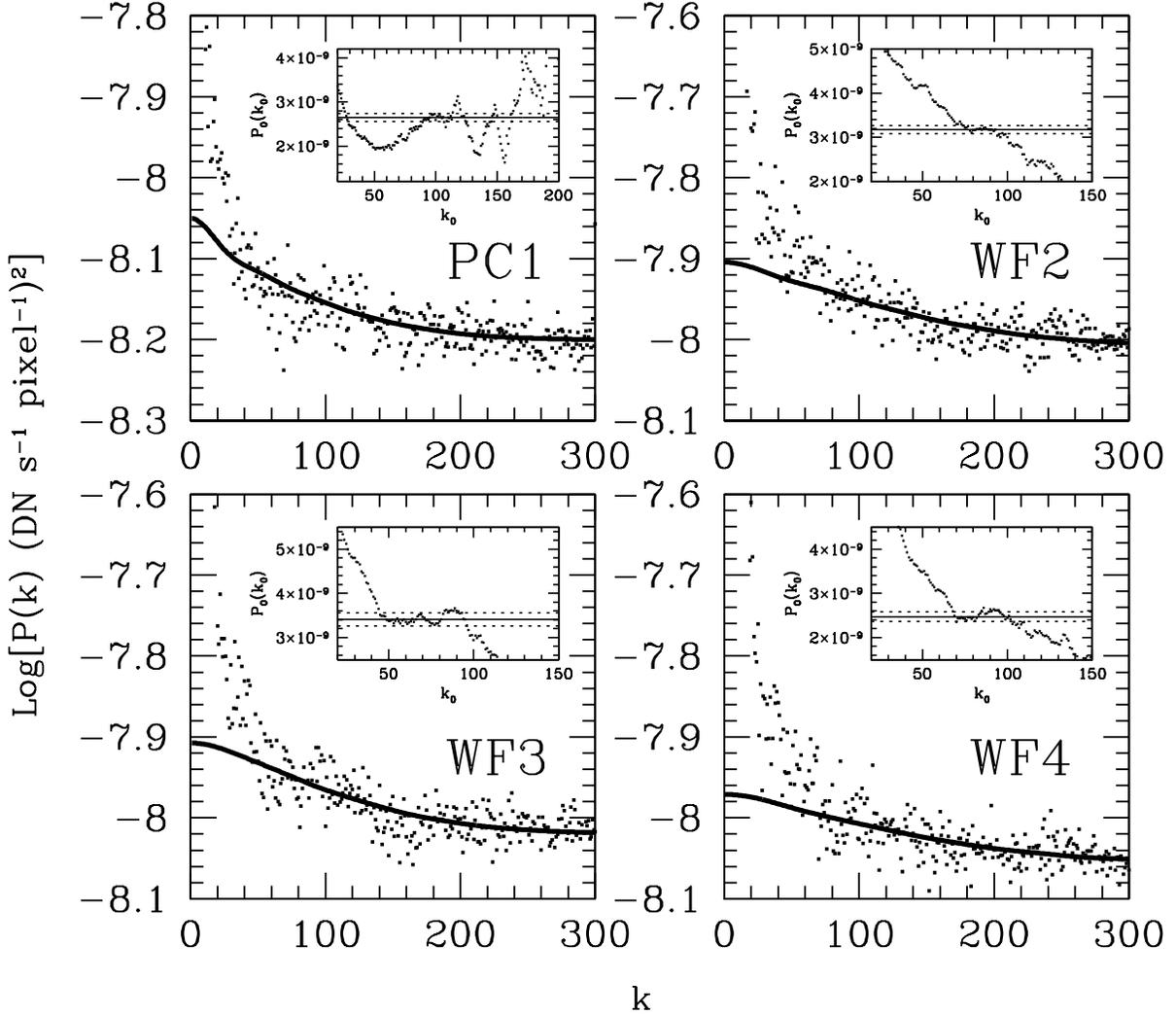}
\caption{Example of a SBF measurement: results for the average image in the F450W filter with $m_c=28.8$. Solid line represents the power spectrum fitting. The observed discrepancy at low wave numbers between the obtained power spectrum and the fit is due to large scale fluctuations in the background brightness of the images. The power-spectrum fitting procedure is the following: eq.~\ref{pow2} is used to fit the power spectrum for wave numbers in the range [$k_0$, $k_{\rm max}$],  $k_{\rm max}$ being the highest wave number of the computed power spectrum, and $k_0$ a number which we vary from 0 to $k_{\rm max}$. As a result, two functions, $P_0(k_0)$ and $P_1(k_0)$, are obtained. The function $P_0(k_0)$ is also shown (small boxes). It can be seen that this function exhibits a ``plateau'' region. The final adopted result and its uncertainty for $P_0$ are obtained computing the average and standard deviation of $P_0(k_0)$ in the plateau interval. For the $P_1$ measurement, the procedure is exactly the same as for $P_0$. \label{$B_{450}$.average.29}}
\end{figure}

\clearpage

\clearpage

\clearpage

\clearpage

\clearpage

\clearpage

\begin{deluxetable}{lcc}
\tabletypesize{\scriptsize}
\tablecaption{The data. \label{t-data}}
\tablewidth{0pt}
\tablehead{
\colhead{Image} & \colhead{Filter} & \colhead{$t_{\rm exp} (s)$}
}
\startdata
F450W.d1.dark  & F450W & 17200 \\
F450W.d2.dark  & F450W & 7900  \\
F450W.d3.dark  & F450W & 11500 \\
F450W.d4.dark  & F450W & 12400 \\
F450W.d5.dark  & F450W & 11800 \\
F450W.d6.dark  & F450W & 12900 \\
F450W.d8.dark  & F450W & 13100 \\
F450W.d9.dark  & F450W & 10700 \\
\\
F606W.d1.dark  & F606W & 6700  \\
F606W.d2.dark  & F606W & 4800  \\
F606W.d3.dark  & F606W & 6450  \\
F606W.d4.dark  & F606W & 10300 \\
F606W.d5.dark  & F606W & 15600 \\
F606W.d6.dark  & F606W & 17300 \\
F606W.d7.dark  & F606W & 14600 \\
F606W.d8.dark  & F606W & 10100 \\
F606W.d9.dark  & F606W & 7100  \\
F606W.d10.dark & F606W & 8300  \\
F606W.d11.dark & F606W & 7800  \\
\\
F814W.d1.dark  & F814W & 10800 \\
F814W.d2.dark  & F814W & 12200 \\
F814W.d3.dark  & F814W & 14200 \\
F814W.d4.dark  & F814W & 12000 \\
F814W.d5.dark  & F814W & 13000 \\
F814W.d6.dark  & F814W & 13100 \\
F814W.d8.dark  & F814W & 5800  \\
F814W.d9.dark  & F814W & 12900 \\
\enddata

\end{deluxetable}

\clearpage
\begin{deluxetable}{lcc}
\tabletypesize{\scriptsize}
\tablecaption{HDF photometric zero points. \label{t-zero}}
\tablewidth{0pt}
\tablehead{
\colhead{Filter} & \colhead{Chip} & \colhead{Magnitude ($AB$ system)}
}
\startdata
F450W & PC1 & 21.92 \\
F450W & WF2 & 21.93 \\
F450W & WF3 & 21.93 \\
F450W & WF4 & 21.90 \\
F606W & PC1 & 23.02 \\
F606W & WF2 & 23.02 \\
F606W & WF3 & 23.03 \\
F606W & WF4 & 23.00 \\
F814W & PC1 & 22.08 \\
F814W & WF2 & 22.09 \\
F814W & WF3 & 22.09 \\
F814W & WF4 & 22.07 \\

\enddata

\end{deluxetable}
\clearpage
\begin{deluxetable}{cccc}
\tabletypesize{\scriptsize}
\tablecaption{$n(m)$-estimated $\sigma_{\rm BG}^2$, in units of $[\frac{DN}{\rm s \times pix}]^2$ \label{t-sigma}}
\tablewidth{0pt}
\tablehead{$m_{\rm c}$ & F450W & F606W & F814W}
\startdata

\multicolumn{4}{c}{From \citet{W96} data} \\
\\

27.8 & 2.82 $\times 10^{-9}$  & 2.74 $\times 10^{-8}$ & 5.50 $\times 10^{-9}$  \\
28.8 & 6.45 $\times 10^{-10}$ & 6.43 $\times 10^{-9}$ & 1.32 $\times 10^{-9}$ \\

\\
\multicolumn{4}{c}{From \citet{Met01} data}\\
\\

27.8 & 7.20 $\times 10^{-9}$  & 5.76 $\times 10^{-8}$ & 1.05 $\times 10^{-8}$  \\
28.8 & 2.23 $\times 10^{-9}$  & 1.62 $\times 10^{-8}$ & 3.04 $\times 10^{-9}$ \\

\enddata

\end{deluxetable}

\clearpage

\begin{deluxetable}{lcc}
\tabletypesize{\scriptsize}
\tablecaption{F450W average image ($m_{\rm c}=27.8$) SBF results varying the patch radius, in units of $10^{-9}\times[\frac{DN}{\rm s \times pix}]^2$ \label{patches}}
\tablewidth{0pt}
\tablehead{
\colhead{Radius (pixels)} & \colhead{$P_0$} & \colhead{$P_1$}
}
\startdata
15 (adopted) & 4.52$\pm$0.06  & 9.01$\pm$0.02  \\
20           & 4.45$\pm$0.05  & 9.05$\pm$0.02  \\
25           & 4.50$\pm$0.06  & 9.03$\pm$0.02  \\
30           & 4.47$\pm$0.06  & 9.00$\pm$0.02  \\

\enddata

\end{deluxetable}
\clearpage

\begin{deluxetable}{l|cc|cc|cc|cc}
\rotate
\tabletypesize{\scriptsize}
\tablecaption{SBF results for $m_{\rm c}=27.8$, in units of $10^{-8}\times[\frac{DN}{\rm s \times pix}]^2$ \label{t-results28}}
\tablewidth{0pt}
\tablehead{
\colhead{Image} &  \multicolumn{2}{c}{\bf{PC1}}&  \multicolumn{2}{c}{\bf{WF2}} & \multicolumn{2}{c}{\bf{WF3}} & \multicolumn{2}{c}{\bf{WF4}} \\
 \colhead{} & \colhead{$P_0$} & \colhead{$P_1$} & \colhead{$P_0$} & \colhead{$P_1$} & \colhead{$P_0$} & \colhead{$P_1$} & \colhead{$P_0$} & \colhead{$P_1$}
}

\startdata
F450W.d1.dark  & 1.90  $\pm$ 0.16 & 3.86 $\pm$ 0.01  & 1.05 $\pm$ 0.11 &  5.4  $\pm$ 0.1  & 0.86  $\pm$ 0.05 & 5.4 $\pm$ 0.1   & 1.09 $\pm$ 0.05 & 4.6   $\pm$ 0.1  \\
F450W.d2.dark  & 3.04  $\pm$ 0.10 & 7.21 $\pm$ 0.01  & 1.89 $\pm$ 0.07 &  11.4 $\pm$ 0.1  & 1.70  $\pm$ 0.23 & 10.8$\pm$ 0.1   & 1.13 $\pm$ 0.17 & 10.0  $\pm$ 0.1  \\
F450W.d3.dark  & 1.48  $\pm$ 0.06 & 5.25 $\pm$ 0.01  & 1.06 $\pm$ 0.10 & 8.0   $\pm$ 0.1  & 1.01  $\pm$ 0.07 & 7.8 $\pm$ 0.1   & 1.33 $\pm$ 0.10 & 6.9   $\pm$ 0.1  \\
F450W.d4.dark  & 1.39  $\pm$ 0.09 & 4.45 $\pm$ 0.01  & 1.34 $\pm$ 0.06 & 6.9   $\pm$ 0.1  & 1.25  $\pm$ 0.09 & 6.7 $\pm$ 0.1   & 1.42 $\pm$ 0.13 & 6.2   $\pm$ 0.1  \\
F450W.d5.dark  & 1.73  $\pm$ 0.09 & 5.18 $\pm$ 0.01  & 1.21 $\pm$ 0.05 & 7.5   $\pm$ 0.1  & 1.36  $\pm$ 0.04 & 7.2 $\pm$ 0.1   & 1.89 $\pm$ 0.07 & 6.3   $\pm$ 0.1  \\
F450W.d6.dark  & 1.47  $\pm$ 0.06 & 4.31 $\pm$ 0.01  & 1.04 $\pm$ 0.12 &  6.9  $\pm$ 0.1  & 1.22  $\pm$ 0.03 & 6.7 $\pm$ 0.1   & 1.00 $\pm$ 0.08 & 6.1   $\pm$ 0.1  \\
F450W.d8.dark  & 1.24  $\pm$ 0.05 & 4.32 $\pm$ 0.01  & 0.77 $\pm$ 0.08 &  6.9  $\pm$ 0.1  & 1.40  $\pm$ 0.05 & 6.7 $\pm$ 0.1   & 0.76 $\pm$ 0.11 & 6.1   $\pm$ 0.1  \\
F450W.d9.dark  & 2.10  $\pm$ 0.08 & 5.95 $\pm$ 0.01  & 1.13 $\pm$ 0.11 &  8.9  $\pm$ 0.1  & 0.98  $\pm$ 0.21 & 8.5 $\pm$ 0.1   & 1.70 $\pm$ 0.06 & 7.7   $\pm$ 0.1  \\
\hline
Average image  &  0.29  $\pm$ 0.03  & 0.636 $\pm$ 0.001  & 0.452$\pm$ 0.006 &  0.901$\pm$ 0.002 & 0.563 $\pm$ 0.014& 0.847$\pm$ 0.004 & 0.441 $\pm$ 0.006 & 0.762  $\pm$ 0.002  \\
\hline
\hline
\\
F606W.d1.dark  & 4.50  $\pm$ 0.24 & 17.6  $\pm$ 0.1  &  7.7  $\pm$ 0.3 & 36.8  $\pm$ 0.1   & 6.8  $\pm$ 0.9 & 37.4 $\pm$ 0.2  & 2.5  $\pm$ 0.2 & 35.2  $\pm$ 0.1  \\
F606W.d2.dark  & 6.1  $\pm$ 0.5  & 27.2  $\pm$ 0.1  &  8.2  $\pm$ 0.2 & 54.9  $\pm$ 0.4   & 5.5  $\pm$ 0.6 & 56.5 $\pm$ 0.1  & 1.4  $\pm$ 0.9 & 52.5  $\pm$ 0.3  \\
F606W.d3.dark  & 4.7  $\pm$ 0.6  & 22.2  $\pm$ 0.1  &  5.5  $\pm$ 0.5 & 42.7  $\pm$ 0.2   & 6.2  $\pm$ 1.7 & 42.1 $\pm$ 0.3  & 8.8  $\pm$ 0.6 & 42.6  $\pm$ 0.3  \\
F606W.d4.dark  & 2.92  $\pm$ 0.21 & 13.3  $\pm$ 0.1  &  4.2  $\pm$ 0.4 & 26.1  $\pm$ 0.1   & 6.2  $\pm$ 0.5 & 25.7 $\pm$ 0.1  & 5.2  $\pm$ 0.4 & 23.9  $\pm$ 0.1  \\
F606W.d5.dark  & 4.71  $\pm$ 0.18 & 9.3   $\pm$ 0.1  &  6.1  $\pm$ 0.2 & 17.6  $\pm$ 0.1   & 6.8  $\pm$ 0.6 & 19.1 $\pm$ 0.2  & 4.5  $\pm$ 0.6 & 16.8  $\pm$ 0.2  \\
F606W.d6.dark  & 2.48  $\pm$ 0.09 & 7.5   $\pm$ 0.1  &  6.5  $\pm$ 0.3 & 16.2  $\pm$ 0.1   & 6.0  $\pm$ 0.2 & 17.4 $\pm$ 0.1  & 5.2  $\pm$ 0.4 & 14.8  $\pm$ 0.1  \\
F606W.d7.dark  & 3.52  $\pm$ 0.16 & 9.0   $\pm$ 0.1  &  6.3  $\pm$ 0.3 & 18.2  $\pm$ 0.1   & 7.1  $\pm$ 0.4 & 18.6 $\pm$ 0.1  & 4.3  $\pm$ 0.2 & 16.7  $\pm$ 0.1  \\
F606W.d8.dark  & 3.6  $\pm$ 0.4  & 12.6  $\pm$ 0.1  &  7.4  $\pm$ 0.5 & 25.2  $\pm$ 0.1   & 6.7  $\pm$ 0.6 & 25.7 $\pm$ 0.1  & 5.6  $\pm$ 0.4 & 24.2  $\pm$ 0.2  \\
F606W.d9.dark  & 5.17  $\pm$ 0.24 & 20.3  $\pm$ 0.1  &  6.6  $\pm$ 0.4 & 37.6  $\pm$ 0.1   & 7.8  $\pm$ 0.3 & 38.4 $\pm$ 0.2  & 6.5  $\pm$ 0.4 & 33.7  $\pm$ 0.2  \\
F606W.d10.dark & 4.50  $\pm$ 0.13 & 15.1  $\pm$ 0.1  &  7.3  $\pm$ 0.4 & 30.2  $\pm$ 0.1   & 7.6  $\pm$ 0.6 & 31.3 $\pm$ 0.2  & 4.7  $\pm$ 0.7 & 28.2  $\pm$ 0.2  \\
F606W.d11.dark & 5.37  $\pm$ 0.16 & 15.5  $\pm$ 0.1  &  5.5  $\pm$ 0.8 & 31.3  $\pm$ 0.2   & 7.4  $\pm$ 0.8 & 32.3 $\pm$ 0.2  & 4.4  $\pm$ 0.7 & 28.9  $\pm$ 0.2  \\
\hline
Average image  & 0.96  $\pm$ 0.18 &  1.38 $\pm$ 0.01 &  3.19 $\pm$ 0.13&  2.68 $\pm$ 0.02  & 3.08 $\pm$ 0.20& 2.52 $\pm$ 0.04 & 3.2 $\pm$ 0.5 & 1.95  $\pm$ 0.14  \\
\hline
\hline
\\
F814W.d1.dark  & 2.24  $\pm$ 0.10 & 7.94  $\pm$ 0.01  & 0.83 $\pm$ 0.16 & 14.9   $\pm$ 0.1  & 1.5   $\pm$ 0.3  & 15.7  $\pm$ 0.1  & 1.8  $\pm$ 0.3  &  14.7 $\pm$ 0.1  \\
F814W.d2.dark  & 1.89  $\pm$ 0.11 & 7.59  $\pm$ 0.01  & 1.99 $\pm$ 0.15 & 13.3   $\pm$ 0.1  & 3.23  $\pm$ 0.09 & 24.4  $\pm$ 0.1  & 1.70 $\pm$ 0.18 &  13.3 $\pm$ 0.1  \\
F814W.d3.dark  & 2.04  $\pm$ 0.04 & 6.13  $\pm$ 0.01  & 1.34 $\pm$ 0.17 & 11.8   $\pm$ 0.1  & 1.68  $\pm$ 0.11 & 11.9  $\pm$ 0.1  & 1.56 $\pm$ 0.07 &  10.5 $\pm$ 0.1  \\
F814W.d4.dark  & 2.03  $\pm$ 0.05 & 6.92  $\pm$ 0.01  & 2.51 $\pm$ 0.23 & 13.0   $\pm$ 0.1  & 1.94  $\pm$ 0.20 & 13.8  $\pm$ 0.1  & 0.75 $\pm$ 0.10 &  13.6 $\pm$ 0.1  \\
F814W.d5.dark  & 2.1  $\pm$ 0.3  & 7.00  $\pm$ 0.01  & 2.6  $\pm$ 0.3  & 12.9   $\pm$ 0.1  & 0.92  $\pm$ 0.14 & 14.6  $\pm$ 0.1  & 2.07 $\pm$ 0.09 &  13.0 $\pm$ 0.1  \\
F814W.d6.dark  & 2.04  $\pm$ 0.11 & 6.62  $\pm$ 0.01  & 2.7  $\pm$ 0.3  & 12.1   $\pm$ 0.1  & 3.09  $\pm$ 0.17 & 13.6  $\pm$ 0.1  & 1.29 $\pm$ 0.16 &  12.8 $\pm$ 0.1  \\
F814W.d8.dark  & 5.98  $\pm$ 0.14 & 20.26 $\pm$ 0.01  & 4.1  $\pm$ 0.3  & 27.5   $\pm$ 0.1  & 3.5   $\pm$ 0.3  & 28.6  $\pm$ 0.1  & 3.0  $\pm$ 0.6  &  25.3 $\pm$ 0.1  \\
F814W.d9.dark  & 2.55  $\pm$ 0.19 & 6.77  $\pm$ 0.01  & 2.0  $\pm$ 0.4  & 12.6   $\pm$ 0.1  & 2.3   $\pm$ 0.3  & 13.0  $\pm$ 0.1  & 2.60 $\pm$ 0.07 &  12.0 $\pm$ 0.1  \\
\hline0
Average image  & 0.413 $\pm$ 0.019& 1.074 $\pm$ 0.002 &1.011 $\pm$ 0.024& 1.669  $\pm$  0.007 & 0.47  $\pm$ 0.05 & 2.057  $\pm$ 0.012 & 0.88 $\pm$ 0.03 &  1.592$\pm$ 0.011  \\
\hline
\enddata

\end{deluxetable}

\clearpage

\begin{deluxetable}{l|cc|cc|cc|cc}
\rotate
\tabletypesize{\scriptsize}
\tablecaption{SBF results for $m_{\rm c}=28.8$, in units of $10^{-8}\times[\frac{DN}{\rm s \times pix}]^2$ \label{t-results29}}
\tablewidth{0pt}
\tablehead{
\colhead{Image} &  \multicolumn{2}{c}{\bf{PC1}}&  \multicolumn{2}{c}{\bf{WF2}} & \multicolumn{2}{c}{\bf{WF3}} & \multicolumn{2}{c}{\bf{WF4}} \\
 \colhead{} & \colhead{$P_0$} & \colhead{$P_1$} & \colhead{$P_0$} & \colhead{$P_1$} & \colhead{$P_0$} & \colhead{$P_1$} & \colhead{$P_0$} & \colhead{$P_1$}
}

\startdata
F450W.d1.dark  & 1.53  $\pm$  0.04 & 3.66 $\pm$ 0.01  & 0.75 $\pm$ 0.08 &  5.4   $\pm$  0.1   & 0.62 $\pm$ 0.06 & 5.4  $\pm$ 0.1   &0.83 $\pm$ 0.04 &  4.7   $\pm$  0.1  \\
F450W.d2.dark  & 3.00  $\pm$  0.09 & 7.14 $\pm$ 0.01  & 2.01 $\pm$ 0.06 & 11.2   $\pm$  0.1   & 1.48 $\pm$ 0.23 & 10.8 $\pm$ 0.1   &1.46 $\pm$ 0.17 &  9.9   $\pm$  0.1  \\
F450W.d3.dark  & 1.43  $\pm$ 0.04  & 5.24 $\pm$ 0.01  & 1.04 $\pm$ 0.11 &  8.0   $\pm$  0.1   & 0.81 $\pm$ 0.07 & 7.8  $\pm$ 0.1   &1.22 $\pm$ 0.05 &  6.9   $\pm$  0.1  \\
F450W.d4.dark  & 1.30  $\pm$ 0.06  & 4.45 $\pm$ 0.01  & 1.09 $\pm$ 0.06 &  6.9   $\pm$  0.1   & 1.14 $\pm$ 0.05 & 6.7  $\pm$ 0.1   &1.23 $\pm$ 0.10 &  6.1   $\pm$  0.1  \\
F450W.d5.dark  & 1.68  $\pm$ 0.08  & 5.09 $\pm$ 0.01  & 0.76 $\pm$ 0.06 &  7.6   $\pm$  0.1   & 1.2  $\pm$ 0.3  & 7.2  $\pm$ 0.1   &1.49 $\pm$ 0.04 &  6.4   $\pm$  0.1  \\
F450W.d6.dark  & 1.46  $\pm$ 0.05  & 4.31 $\pm$ 0.01  & 0.59 $\pm$ 0.04 &  6.9   $\pm$  0.1   & 0.67 $\pm$ 0.18 & 6.8  $\pm$ 0.1   &0.77 $\pm$ 0.06 &  6.1   $\pm$  0.1  \\
F450W.d8.dark  & 1.55  $\pm$ 0.04  & 4.29 $\pm$ 0.01  & 0.71 $\pm$ 0.07 &  6.9   $\pm$  0.1   & 1.16 $\pm$ 0.07 & 6.8  $\pm$ 0.1   &0.81 $\pm$ 0.07 &  6.0   $\pm$  0.1  \\
F450W.d9.dark  & 1.97  $\pm$ 0.06  & 5.96 $\pm$  0.01 & 1.32 $\pm$ 0.03 &  8.8   $\pm$  0.1   & 1.2  $\pm$ 0.4  & 8.4  $\pm$ 0.1   &1.57 $\pm$ 0.05 &  7.7   $\pm$  0.1  \\
\hline
Average image  & 0.265 $\pm$ 0.009  & 0.626  $\pm$ 0.001 & 0.317$\pm$ 0.009&  0.929 $\pm$  0.003 & 0.341$\pm$ 0.015& 0.896$\pm$ 0.004 &0.247 $\pm$ 0.011 &  0.822  $\pm$  0.004  \\
\hline
\hline
\\
F606W.d1.dark  & 4.75  $\pm$ 0.17  & 17.7  $\pm$ 0.1  & 5.8  $\pm$ 0.3  & 37.2 $\pm$ 0.1  & 6.0  $\pm$ 0.6  & 37.7 $\pm$ 0.2  & 2.4  $\pm$ 0.7 & 35.0 $\pm$ 0.3  \\
F606W.d2.dark  & 5.9   $\pm$ 0.8   & 27.3  $\pm$ 0.1  & 8.9  $\pm$ 0.8  & 55.2 $\pm$ 0.3  & 4.2  $\pm$ 0.7  & 56.8 $\pm$ 0.2  & 3.6  $\pm$ 0.7 & 51.9 $\pm$ 0.3  \\
F606W.d3.dark  & 4.9   $\pm$ 0.4   & 21.2  $\pm$ 0.1  & 3.8  $\pm$ 0.4  & 43.0 $\pm$ 0.1  & 4.7  $\pm$ 1.7  & 42.5 $\pm$ 0.5  & 6.0  $\pm$ 0.8 & 44.8 $\pm$ 0.3  \\
F606W.d4.dark  & 2.49  $\pm$ 0.22  & 13.4  $\pm$ 0.1  & 1.9  $\pm$ 0.4  & 26.5 $\pm$ 0.1  & 4.4  $\pm$ 0.5  & 26.1 $\pm$ 0.2  & 4.2  $\pm$ 0.4 & 24.2 $\pm$ 0.1  \\
F606W.d5.dark  & 4.69  $\pm$ 0.17  & 9.3   $\pm$ 0.1  & 4.0  $\pm$ 0.2  & 18.0 $\pm$ 0.1  & 4.4  $\pm$ 0.3  & 19.1 $\pm$ 0.1  & 3.2  $\pm$ 0.4 & 16.9 $\pm$ 0.2  \\
F606W.d6.dark  & 2.41  $\pm$ 0.12  & 7.6   $\pm$ 0.1  & 4.2  $\pm$ 0.2  & 16.6 $\pm$ 0.1  & 4.1  $\pm$ 0.2  & 17.6 $\pm$ 0.4  & 3.8  $\pm$ 0.4 & 15.1 $\pm$ 0.2  \\
F606W.d7.dark  & 2.9   $\pm$ 0.4   & 9.1   $\pm$ 0.1  & 4.4  $\pm$ 0.3  & 18.6 $\pm$ 0.1  & 4.5  $\pm$ 0.3  & 19.2 $\pm$ 0.1  & 3.8  $\pm$ 0.7 & 16.8 $\pm$ 0.3  \\
F606W.d8.dark  & 3.48  $\pm$ 0.16  & 12.6  $\pm$ 0.1  & 4.9  $\pm$ 0.5  & 25.8 $\pm$ 0.2  & 5.5  $\pm$ 1.0  & 26.1 $\pm$ 0.3  & 2.5  $\pm$ 1.3 & 24.3 $\pm$ 0.4  \\
F606W.d9.dark  & 5.3   $\pm$ 0.3   & 20.6  $\pm$ 0.1  & 4.9  $\pm$ 0.6  & 37.6 $\pm$ 0.2  & 4.4  $\pm$ 0.4  & 39.2 $\pm$ 0.1  & 4.8  $\pm$ 0.4 & 34.1 $\pm$ 0.2  \\
F606W.d10.dark & 4.41  $\pm$ 0.20  & 15.0  $\pm$ 0.1  & 6.2  $\pm$ 0.3  & 30.5 $\pm$ 0.1  & 6.4  $\pm$ 0.8  & 31.7 $\pm$ 0.2  & 2.6  $\pm$ 0.7 & 28.7 $\pm$ 0.3  \\
F606W.d11.dark & 5.6   $\pm$ 0.4   & 15.4  $\pm$ 0.1  & 3.4  $\pm$ 1.0  & 31.8 $\pm$ 0.3  & 5.3  $\pm$ 0.4  & 33.1 $\pm$ 0.1  & 2.3  $\pm$ 0.5 & 29.2 $\pm$ 0.2  \\
\hline
Average image  &  0.557$\pm$ 0.021 & 1.41  $\pm$ 0.01 & 1.38 $\pm$ 0.15 & 2.81 $\pm$ 0.02 & 1.72 $\pm$ 0.10 & 2.71 $\pm$ 0.02 & 1.4  $\pm$ 0.3 & 2.43 $\pm$ 0.09 \\
\hline
\hline
\\
F814W.d1.dark  & 2.34  $\pm$ 0.08  & 7.99  $\pm$  0.01  & 0.28 $\pm$ 0.12 & 15.1  $\pm$ 0.1  & 1.3  $\pm$ 0.3  & 15.9  $\pm$ 0.1 & 2.02 $\pm$ 0.23 & 14.8 $\pm$ 0.1 \\
F814W.d2.dark  & 1.51  $\pm$ 0.09  & 7.64  $\pm$  0.01  & 1.26 $\pm$ 0.13 & 13.3  $\pm$ 0.1  & 3.6  $\pm$ 0.5  & 26.2  $\pm$ 0.1 & 2.09 $\pm$ 0.17 & 13.2 $\pm$ 0.1 \\
F814W.d3.dark  & 2.04  $\pm$ 0.04  & 6.10  $\pm$  0.01  & 1.43 $\pm$ 0.09 & 11.9  $\pm$ 0.1  & 1.54 $\pm$ 0.22 & 12.0  $\pm$ 0.1 & 1.00 $\pm$ 0.09 & 10.6 $\pm$ 0.1 \\
F814W.d4.dark  & 2.14  $\pm$ 0.05  & 6.91  $\pm$  0.01  & 1.48 $\pm$ 0.08 & 13.3  $\pm$ 0.1  & 1.54 $\pm$ 0.11 & 13.9  $\pm$ 0.1 & 0.9  $\pm$ 0.4  & 13.9 $\pm$ 0.1 \\
F814W.d5.dark  & 2.1   $\pm$ 0.3   & 7.02  $\pm$  0.01  & 1.4  $\pm$ 0.5  & 13.1  $\pm$ 0.1  & 1.37 $\pm$ 0.16 & 14.9  $\pm$ 0.1 & 1.6  $\pm$ 0.3  & 13.2 $\pm$ 0.1 \\
F814W.d6.dark  & 1.79  $\pm$ 0.06  & 6.67  $\pm$  0.01  & 2.12 $\pm$ 0.12 & 12.5  $\pm$ 0.1  & 2.83 $\pm$ 0.15 & 13.2  $\pm$ 0.1 & 1.95 $\pm$ 0.12 & 12.9 $\pm$ 0.1 \\
F814W.d8.dark  & 5.2   $\pm$ 0.5   & 20.81 $\pm$  0.01  & 4.27 $\pm$ 0.15 & 27.7  $\pm$ 0.1  & 3.2  $\pm$ 0.3  & 28.6  $\pm$ 0.1 & 2.3  $\pm$ 0.3  & 25.4 $\pm$ 0.1 \\
F814W.d9.dark  & 2.61  $\pm$ 0.18  & 6.78  $\pm$  0.01  & 2.13 $\pm$ 0.12 & 12.3  $\pm$ 0.1  & 2.50 $\pm$ 0.06 & 13.0  $\pm$ 0.1 & 2.04 $\pm$ 0.07 & 12.3 $\pm$ 0.1 \\
\hline
Average image  & 0.397 $\pm$ 0.022 & 1.084 $\pm$  0.002 & 0.54 $\pm$ 0.03 & 1.787 $\pm$ 0.007& 0.484$\pm$0.022 & 2.140 $\pm$0.006& 0.43 $\pm$ 0.03 & 1.763$\pm$ 0.010\\
\hline
\enddata

\end{deluxetable}

\clearpage

\begin{deluxetable}{lcc|cc|cc}
\tabletypesize{\scriptsize}
\tablecaption{SBF-measured $\sigma_{\rm BG}^2$ and $\sigma_{\rm CR}^2$ results, in units of $10^{-8}\times[\frac{DN}{\rm s \times pix}]^2$ \label{t-results-bg}}
\tablewidth{0pt}
\tablehead{
\colhead{} & \multicolumn{2}{c}{\bf{F450W}}&  \multicolumn{2}{c}{\bf{F606W}} & \multicolumn{2}{c}{\bf{F814W}} \\
\colhead{Chip} & \colhead{$\sigma_{\rm BG}^2$} & \colhead{$\sigma_{\rm cr}^2$} & \colhead{$\sigma_{\rm BG}^2$} & \colhead{$\sigma_{\rm cr}^2$} & \colhead{$\sigma_{\rm BG}^2$} & \colhead{$\sigma_{\rm cr}^2$}
}

\startdata
\multicolumn{7}{c}{$m_{\rm c}=27.8$} \\
\hline
PC1\tablenotemark{(a)} &  0.35   $\pm$ 0.16   &  0.215 $\pm$ 0.007  &  3.1  $\pm$  0.9    & 0.336  $\pm$  0.020  & 0.48   $\pm$  0.10   & 0.314 $\pm$ 0.006  \\
WF2                    &  0.347  $\pm$ 0.008  &  0.105 $\pm$ 0.005  &  2.26 $\pm$  0.13   & 0.331  $\pm$  0.018  & 0.83   $\pm$  0.03   & 0.176 $\pm$ 0.012  \\
WF3                    &  0.468  $\pm$ 0.015  &  0.094 $\pm$ 0.006  &  2.72 $\pm$  0.20   & 0.36   $\pm$  0.03   & 0.52   $\pm$  0.05   & 0.218 $\pm$ 0.013  \\
WF4                    &  0.320  $\pm$ 0.008  &  0.121 $\pm$ 0.005  &  3.0  $\pm$  0.5    & 0.16   $\pm$  0.05   & 0.74   $\pm$  0.04   & 0.139 $\pm$ 0.021  \\
SBF-measured $ \sigma_{\rm BG}^2$  &  0.37   $\pm$ 0.04   &                     &  2.8  $\pm$  0.3    &                      & 0.64   $\pm$  0.03   &           \\
\hline
\multicolumn{7}{c}{$m_{\rm c}=28.8$} \\
\hline
PC1\tablenotemark{(a)} & 0.26    $\pm$  0.05   &  0.211  $\pm$  0.006  & 0.91  $\pm$  0.11  &  0.370 $\pm$  0.011  & 0.42  $\pm$  0.12   & 0.297 $\pm$ 0.012   \\
WF2                    & 0.214   $\pm$  0.010  &  0.102  $\pm$  0.004  & 1.04  $\pm$  0.15  &  0.342 $\pm$  0.021  & 0.36  $\pm$  0.03   & 0.179 $\pm$ 0.011   \\
WF3                    & 0.242   $\pm$  0.019  &  0.099  $\pm$  0.011  & 1.40  $\pm$  0.10  &  0.32  $\pm$  0.03   & 0.23  $\pm$  0.03   & 0.251 $\pm$ 0.013   \\
WF4                    & 0.115   $\pm$  0.012  &  0.132  $\pm$  0.004  & 1.2   $\pm$  0.3   &  0.22  $\pm$  0.04   & 0.24  $\pm$  0.03   & 0.187 $\pm$ 0.013   \\
SBF-measured $ \sigma_{\rm BG}^2$   & 0.208   $\pm$  0.014  &                       & 1.13  $\pm$  0.09  &                      & 0.31  $\pm$  0.03   &       \\

\enddata
\tablenotetext{a}{Results from the planetary camera have been scaled to the wide field pixel size.}
\end{deluxetable}

\clearpage

\begin{deluxetable}{lcc}
\tabletypesize{\scriptsize}
\tablecaption{Computed and observed $P_1$ for F450W (WF2) and $m_{\rm c}=28.8$, in units of $10^{-8}\times[\frac{DN}{\rm s \times pix}]^2$ \label{t-p1}}
\tablewidth{0pt}
\tablehead{
\colhead{Image} & \colhead{Computed $P_1$} & \colhead{Observed $P_1$}
}
\startdata
F450W.d1.dark & 5.2  &  5.48  $\pm$  0.02\\
F450W.d2.dark & 11.5 & 11.25  $\pm$  0.01\\
F450W.d3.dark & 7.8  &  8.02  $\pm$  0.03\\
F450W.d4.dark & 6.9  &  6.96  $\pm$  0.02\\
F450W.d5.dark & 7.3  &  7.62  $\pm$  0.02\\
F450W.d6.dark & 6.7  &  6.99  $\pm$  0.01\\
F450W.d8.dark & 6.6  &  6.91  $\pm$  0.02\\
F450W.d9.dark & 8.3  &  8.81  $\pm$  0.01\\
\enddata

\end{deluxetable}

\clearpage

\begin{deluxetable}{lccc}
\tabletypesize{\scriptsize}
\tablecaption{$n(m)$-estimated and SBF-measured $[\sigma_{\rm BG}^2]$, in units of $10^{-8}\times[\frac{DN}{\rm s \times pix}]^2$ \label{t-interval}}
\tablewidth{0pt}
\tablehead{
\colhead{Filter} & \colhead{$n(m)$-estimated} & \colhead{$n(m)$-estimated} & \colhead{SBF-measured} \\
 &using \citet{W96}  & using \citet{Met01}&
}
\startdata
F450W & 0.22   & 0.50 & 0.16       $\pm$  0.04 \\
F606W & 2.10   & 4.14 & 1.6        $\pm$  0.3  \\
F814W & 0.42   & 0.75 & 0.33       $\pm$  0.04 \\
\enddata

\end{deluxetable}

\clearpage

\begin{deluxetable}{ccc}
\tabletypesize{\scriptsize}
\tablecaption{$n(m)$ Results \label{t-GLF}}
\tablewidth{0pt}
\tablehead{
\colhead{Filter} &  $\gamma$ & SBF limiting magnitude
}
\startdata
F450W     &  0.27    &  31.0   \\
F606W     &  0.21    &  30.7   \\
F814W     &  0.26    &  30.8   \\

\enddata

\end{deluxetable}

\end{document}